\documentclass[twocolumn,aps,a4paper,superscriptaddress]{revtex4-2}
\usepackage{kpfonts}
\usepackage{amsmath}
\usepackage{amssymb}
\usepackage{times}
\usepackage{graphicx}
\usepackage[colorlinks,linkcolor=blue,citecolor=blue,urlcolor=blue,breaklinks]{hyperref}
\usepackage[normalem]{ulem}
\newcommand{\be}{\begin{equation}}
\newcommand{\ee}{\end{equation}}
\newcommand{\bea}{\begin{eqnarray}}
\newcommand{\eea}{\end{eqnarray}}

\def\b{\beta}

\def\G{\Gamma}
\def\d{\delta}
\def\D{\Delta}
\def\e{\epsilon}

\def\th{\theta}

\def\l{\lambda}

\def\m{\mu}
\def\n{\nu}

\def\p{\pi}

\def\r{\rho}
\def\s{\sigma}
\def\S{\Sigma}
\def\t{\tau}

\def\w{\omega}

\def\Q{\Psi}

\def\blk{{\mathbf k}}

\def\blQ{{\mathbf Q}}

\def\callE{\mbox{$\mathcal{E}$}}

\def\callR{\mbox{$\mathcal{R}$}}

\def\bra{\langle}
\def\ket{\rangle}

\def\Tr{{\rm Tr}}

\def\1op{\hat{\mathbbm{1}}}
\def\nn{\nonumber}

\def\bz{\mathbf 0}

\begin{document}

\title{Unified First-Principles Theory of Coherent and Incoherent Excitons in Time-Resolved ARPES}
\title{Unified First-Principles Formula for Time-Resolved ARPES 
Spectra of Coherent and Incoherent Excitons}
\author{Gianluca Stefanucci}
\affiliation{Dipartimento di Fisica, Universit{\`a} di Roma Tor Vergata, Via della Ricerca Scientifica 1,
00133 Rome, Italy}
\affiliation{INFN, Sezione di Roma Tor Vergata, Via della Ricerca Scientifica 1, 00133 Rome, Italy}

\author{Enrico Perfetto}
\affiliation{Dipartimento di Fisica, Universit{\`a} di Roma Tor Vergata, Via della Ricerca Scientifica 1,
00133 Rome, Italy}
\affiliation{INFN, Sezione di Roma Tor Vergata, Via della Ricerca Scientifica 1, 00133 Rome, Italy}

\begin{abstract}  	
Despite major experimental progresses in time-resolved 
and angle-resolved photoemission spectroscopy, a  quantitative,
microscopic framework for interpreting 
exciton-induced modifications of electronic band structures -- 
applicable even beyond the low-density limit -- is still lacking. 
Here we close this gap by introducing a unified 
approach that links the dynamics of 
coherent and incoherent excitons to distinct and experimentally observable
excitonic sidebands. 
Our central result is a general, first-principles 
formula for time-resolved photoemission spectra, applicable across a 
broad range of temperatures, excitation densities, and pump–probe 
delays. This advance provides a predictive tool for 
quantitatively tracking excitonic dynamics in complex 
materials.      
\end{abstract}

\maketitle

{\em Introduction.--}
The concept of excitons as quasiparticles in semiconductors and 
insulators has long been central to the interpretation of optical 
spectra, where it underpins the understanding of subgap resonances 
and above-gap structures in absorbance and reflectance 
measurements~\cite{onida_electronic_2002}. Optical probes, however, 
are fundamentally restricted to bright, zero–center-of-mass-momentum 
excitons and primarily access their energies. By contrast, 
angle-resolved photoemission spectroscopy (ARPES) reveals
a far 
richer and more detailed 
picture~\cite{zhang_angle-resolved_2022,boschini_time-resolved_2024,xu_time-domain_2025}. 
In particular, time-resolved ARPES (TR-ARPES) 
has recently 
emerged as a powerful platform for exciton physics, enabling  direct access 
to bright, momentum-dark, and spin-dark 
excitons~\cite{madeo_directly_2020,dendzik_observation_2020,tanimura_time_2020,wallauer_momentum-resolved_2021,fukutani_detecting_2021,kunin_momentum-resolved_2023,bange_probing_2024,gosetti_unveiling_2024}. 
Beyond exciton energies, TR-ARPES uniquely resolves excitonic 
wavefunctions in both band and momentum 
space~\cite{man_experimental_2021,dong_direct_2021,meneghini_hybrid_2023,mori_spin-polarized_2023}.

Although TR-ARPES~\cite{na_advancing_2023} has led to 
informative experimental spectra,  a rigourous 
connection with the underlying exciton dynamics  is still 
lacking.  
After pumping below the bandgap,  
the nonequilibrium exciton 
fluid~\cite{haug2009quantum,lindberg_effective_1988,ostreich_non-stationary_1993,hannewald_excitonic_2000,koch_semiconductor_2006} 
drives the motion of the nuclear 
lattice. 
The scattering between electrons and phonons 
destroys exciton 
coherence~\cite{madrid_phonon-induced_2009,nie_ultrafast_2014,dey_optical_2016,sangalli_an-ab-initio_2018,stefanucci_from-carriers_2021,perfetto_real-time_2023} 
and diffuses 
excitons~\cite{bertoni_generation_2016,uddin_neutral_2020,perfetto_ultrafast_2021,stefanucci_excitonic_2025}. 
Bright coherent excitons are converted 
-- typically in less than a few hundreds of femtoseconds~\cite{moody_intrinsic_2015,selig_exciton_2016,jakubczyk_impact_2018} -- into 
bright and dark incoherent excitons.
TR-ARPES spectra of {\em coherent} excitons have been successfully 
described~\cite{perfetto_pump-driven_2019,perfetto_time-resolved_2020,perfetto_floquet_2020,chan_giant_2023,pareek_driving_2026}.
However,  a first principles many-body framework for 
the {\em incoherent} regime -- applicable even beyond the 
domain of low excitation densities -- remains underdeveloped~\cite{perfetto_first-principles_2016,steinhoff_exciton_2017,rustagi_photoemission_2018,christiansen_theory_2019}. 
In this Letter, we identify distinct signatures of coherent and 
incoherent excitons in the TR-ARPES signal and provide a unified 
theory of excitonic sidebands emerging from different exciton states.  
The main finding is a 
ready-to-use first-principles formula for interpreting TR-ARPES spectra  
across a wide range 
of temperatures, excitation densities and pump-probe delays.

{\em Photocurrent.--}
The rate of photoelectrons with energy $\e$ and parallel 
momentum $\blk$, ejected by a high-energy probe 
pulse $a(t)=ae^{-i\w_{0}t}+{\rm h.c.}$ 
impinging on the crystal after a delay $\t$ from the pump, is 
proportional to~\cite{freericks_theoretical_2009,perfetto_first-principles_2016}
\begin{align}
I_{\blk}(\t,\e)&\propto-i|a|^{2}\sum_{\m\m'}
D_{\m\blk}(\e)D^{\ast}_{\m'\blk}(\e)
G^{<}_{\m\m'\blk}(\t,\w_{0}-\e),
\label{tdph2}
\end{align}
where  $D_{\m\blk}(\e)$ is the 
photoemission matrix element for an electron
with band-momentum index $(\m\blk)$. 
Henceforth, we use $\m=c$ for conduction bands and $\m=v$ for valence 
bands. The nonequilibrium electronic 
properties of the crystal are encoded in the Fourier transform, 
$G^{<}(\t,\w)$,  
of the lesser Green's function $G^{<}(t,t')$ with respect to the 
time-difference $t-t'$ for a given center-of-mass-time (delay) $\t=(t+t')/2$.
In the quasiparticle approximation 
$G^{<}_{\m\m'\blk}(\t,\w)=\d_{\m\m'}2\p i 
f_{\m\blk}(\t)\d(\w-\e_{\m\blk}(\t))$, and the spectrum is peaked at 
the quasiparticle energies $\e_{\m\blk}(\t)$ -- with an intensity 
proportional to the 
nonequilibrium populations $f_{\m\blk}(\t)$.
Since our primary interest is in the excitonic sidebands appearing 
above the valence band maximum, 
we  focus exclusively on the contribution of conduction electrons  to  
the photoemission signal. Consequently, we restrict
the sum over $\m,\m'$ in Eq.~(\ref{tdph2}) to the conduction bands.

{\em TR-ARPES formula.--}    
The interacting 
nonequilibrium $G^{<}$ for conduction electrons can be calculated 
from~\cite{svl-book_2025} 
\begin{align}
G^{<}_{\blk}(\t,\w)=G^{R}_{\blk}(\t,\w)\S^{<}_{\blk}(\t,\w)G^{A}_{\blk}(\t,\w),
\label{G<}
\end{align}
where all quantities are matrices with conduction band indices. 
We remark that this way of calculating $G^{<}$ 
differs from  
earlier
attempts~\cite{perfetto_first-principles_2016,steinhoff_exciton_2017}, which 
rely on the fluctuation-dissipation {\em ansatz} 
$G^{<}_{\blk}(\t,\w)=i f(\w-\m_{c})A_{\blk}(\t,\w)$, where $f$ is the Fermi 
function with a conduction-band
chemical potential $\m_{c}$ fixed by the excitation density,
and $A_{\blk}(\t,\w)$ is the nonequilibrium spectral function. This 
ansatz is problematic for frequencies $\w$ below the gap since, at finite 
temperature and/or out of 
equilibrium, the addition and removal energies 
overlap.
The retarded and advanced Green's functions 
\begin{align}
G^{R}_{\blk}(\t,\w)=[G^{A}_{\blk}(\t,\w)]^{\dag}=[\w-\callE_{\blk}(\t)-\S^{R}_{\blk}(\t,\w)],
\end{align}
with $[\callE_{\blk}(\t)]_{\m\m'}\equiv \d_{\m\m'}\e_{\m\blk}(\t)$, 
and the lesser 
Green's function in Eq.~(\ref{G<}) can be calculated once an approximation for the 
electronic self-energy $\S$ is available.

Let us consider  pump fields with frequencies 
below the bandgap, ensuring that only discrete states are 
explored during the time evolution.
We demonstrate below that Hartree plus statically 
screened exchange (HSEX) effectively capture 
the coherent regime, while T-matrix plus exchange 
accurately describes the incoherent regime. 
Remarkably, the full self-energy can be compactly 
expressed as 
\begin{subequations}
\begin{align}
\S^{R}_{cc'\blk}(\t,\w)&=\sum_{\l\blQ v}\s^{\l\blQ 
v}_{cc'\blk}(\t,\w)\;,
\label{sigmaR}
\\
\S^{<}_{cc'\blk}(\t,\w)&=\sum_{\l\blQ v}
f(\w-E_{\l\blQ}(\t)-\e_{v\blk-\blQ}(\t))
\nn\\
&\times[\s^{\l\blQ v}_{cc'\blk}(\t,\w)-\s^{\l\blQ 
v\ast}_{cc'\blk}(\t,\w)]\;,
\label{sigma<}
\end{align}
\label{sigma}
\end{subequations}
where the sum runs over all valence bands and over all exciton states $\l$ of 
momentum $\blQ$. In the Fermi function of Eq.~(\ref{sigma<}), 
the quasiparticle energies 
$\e_{\m\blk}(\t)$ and exciton energies $E_{\l\blQ}(\t)$ are those of 
the {\em nonequilibrium} system.
The key quantity in Eq.~(\ref{sigma}) is the function
\begin{align}
\s^{\l\blQ v}_{cc'\blk}(\t,\w)=K^{\l\blQ}_{cv\blk}\;
\frac{N_{\l\blQ}(\t)}{\w-E_{\l\blQ}(\t)-\e_{v\blk-\blQ}(\t)+i\eta}\;
K^{\l\blQ\ast}_{c'v\blk}-i\eta\;,
\label{sigmasmall}
\end{align}
where $\eta$ is a positive infinitesimal, and the amplitude
\begin{align}
K^{\l\blQ}_{cv\blk}=(\e^{\rm eq}_{c\blk}-\e^{\rm eq}_{v\blk-\blQ}-E^{\rm eq}_{\l\blQ})A^{\l 
\blQ}_{cv\blk-\blQ}
\label{contrdef}
\end{align}
is expressed in terms of the 
equilibrium band structure $\e^{\rm eq}_{\m\blk}$, 
exciton energies $E^{\rm eq}_{\l\blQ}$ and wavefunctions $A^{\l 
\blQ}_{cv\blk}$. Notably, the full self-energy  
depends only on the {\em total} number of 
excitons at the probing time $\t$~\cite{stefanucci_excitonic_2025}
\begin{align}
N_{\l\blQ}(\t)=\d_{\blQ,\bz}|\r_{\l}(\t)|^{2}+
N^{\rm inc}_{\l\blQ}(\t),
\label{N=c+ixb}
\end{align}
with $|\r_{\l}|^{2}$ the number of coherent excitons,
and $N^{\rm inc}_{\l\blQ}$ the number of incoherent 
excitons. 
In Appendix~\ref{findensapp} we present a 
practical method to extract the nonequilibrium energies
$\e_{\m\blk}(\t)$ and $E_{\l\blQ}(\t)$ 
directly from $N_{\l\blQ}(\t)$. 
Equations~(\ref{G<}-\ref{sigmasmall}) are the main 
result of this Letter. They 
 provide a first-principles formula 
to calculate TR-ARPES spectra for any pump-probe delay, and for a 
wide range of temperatures and excitation densities.
The only input quantities are
the exciton 
populations $N_{\l\blQ}(\t)$, which
can be determined by, e.g., solving the excitonic Bloch 
equations~\cite{thranhardt_quantum_2000,selig_exciton_2016,brem_exciton_2018,stefanucci_excitonic_2025}.

{\em Coherent regime.--} Here we briefly revisit the 
theory of the coherent regime, and offer an alternative perspective.
Without any 
loss of generality we set 
the time origin, $t=0$, after the action of the pump pulse. For weak 
pumping the excitation density is low, and 
the state of the system at positive times is given by the quantum 
superposition  
$|\Q(t)\ket=|\Q_{g}\ket+\sum_{\l}\r_{\l}(t)|\l\bz\ket$, where 
$|\Q_{g}\ket$ is the ground state and 
$|\l\blQ\ket=\sum_{cv\blk}A^{\l\blQ}_{cv\blk}\hat{d}^{\dag}_{c\blk+\blQ}\hat{d}_{v\blk}|\Q_{g}\ket$
are the exciton states of momentum $\blQ$ -- here 
$\hat{d}_{\m\blk}$ is the annihilation operator for electrons in band 
$\m$ with momentum $\blk$. As the pump pulse 
can only excite bright excitons,  $|\Q(t)\ket$ contains 
exclusively excitons of vanishing momentum, $\blQ=\bz$. The coefficients 
$\r_{\l}(t)=\sum_{cv\blk}A^{\l\bz\ast}_{cv\blk}\,
\bra\Q(t)|\hat{d}^{\dag}_{v\blk}\hat{d}_{c\blk}|\Q(t)\ket$ are the 
exciton polarizations, and their square modulus yield the number of 
coherent excitons, see Eq.~(\ref{N=c+ixb}). The
system is described by the state $|\Q(t)\ket$  
up to times smaller than the exciton lifetimes.
Such time window defines the coherent regime. From 
the knowledge of $|\Q(t)\ket$ we can directly calculate $G^{<}$, see 
Appendix~\ref{lehmannapp}, and hence the 
TR-ARPES spectrum. We refer to this strategy as the {\em Lehmann 
approach}, since it is based on expanding the many-body state in 
eigenstates of the Hamiltonian.
The same $G^{<}$ can be 
obtained from Eq.~(\ref{G<}) using 
the HSEX self-energy, see Eq.~(\ref{G<tmatfinld}). 
The explicit form 
of the HSEX self-energy can be readily derived
by projecting the equations of motion for the Green's 
function~\cite{perfetto_pump-driven_2019,perfetto_time-resolved_2020} onto 
the conduction-band subspace. The final result is 
Eqs.~(\ref{sigma}) with $\s$ evaluated in 
$N_{\l\blQ}=\d_{\blQ\bz}|\r_{\l}|^{2}$, see 
Appendices~\ref{hsextwobandapp} and~\ref{hsexmultbandapp} for details.
The advantage of the many-body formulation over the 
Lehmann approach is 
that it successfully extends beyond the domain of low excitation 
densities~\cite{pareek_driving_2026}. 

{\em Incoherent regime.--}
In the incoherent regime the 
polarization $\r_{\l}$ vanishes, and consequently the HSEX 
self-energy in Eqs.~(\ref{sigma}) vanishes too. 
In this regime, the 
electronic subsystem is described by an ensemble of bright and dark 
incoherent excitons.
The transition from a pure state to an ensemble is a general 
phenomenon in systems interacting with a bath, such as  
electrons interacting with phonons in our case.    
It has been demonstrated in an exactly  solvable two-band 
model~\cite{stefanucci_from-carriers_2021,zhang_long-lived_2025}, 
and it is also a consequence of the excitonic Bloch 
equations~\cite{thranhardt_quantum_2000,selig_exciton_2016,brem_exciton_2018,stefanucci_excitonic_2025}. 

A heuristic theory for the photocurrent in the incoherent regime 
can be effectively developed for low excitation densities. 
According to the excitonic Bloch equations the incoherent ensemble is described 
by the many-body density matrix $\hat{\r}=|\Q_{g}\ket\bra\Q_{g}|+\sum_{\l\blQ}N^{\rm 
inc}_{\l\blQ}|\l\blQ\ket\bra\l\blQ|$. From the knowledge of 
$\hat{\r}$ we can calculate $G^{<}$, see Appendix~\ref{lehmannapp}, 
which in turn allows us to determine
the TR-ARPES spectrum. This is the Lehmann approach of the incoherent 
regime~\cite{perfetto_first-principles_2016,sangalli_an-ab-initio_2018,rustagi_photoemission_2018,stefanucci_from-carriers_2021,chen_first-principles_2022}. 
Establishing the many-body diagrammatic approximation
underlying this approach 
would lift the low-density constraint, and pave the way toward 
a fully first-principles treatment of TR-ARPES spectra in the 
incoherent  regime.      

At present, the state-of-the-art many-body
approximation is largely based on 
the T-matrix diagrams~\cite{kremp_equation_1984,semkat_ionization_2009,perfetto_first-principles_2016,steinhoff_exciton_2017}.
Previous studies reported 
excitonic sidebands appearing as replicas 
of the conduction bands~\cite{perfetto_first-principles_2016,steinhoff_exciton_2017}, 
in apparent contradiction to the Lehmann 
approach, which instead predicts replicas of the valence bands.   
Another issue pertains to the intensity of the photoemission signal, 
proportional to $|A^{\l\blQ}_{cv\blk}|^{2}$. The 
exciton wavefunctions satisfy the Bethe 
Salpeter equation (BSE), whose
kernel is the sum of the bare 
Coulomb interaction $v$ (direct term) and the screened interaction $W$ 
(exchange term)~\cite{onida_electronic_2002,strinati_application_1988,svl-book_2025}, i.e.,
\begin{align}
(\e^{\rm eq}_{c\blk+\blQ}-\e^{\rm eq}_{v\blk})A^{\l\blQ}_{cv\blk}-
\sum_{c'v'\blk'}
K^{\blQ}_{cv\blk,c'v'\blk'}A^{\l\blQ}_{c'v'\blk'}=
E^{\rm eq}_{\l\blQ}A^{\l\blQ}_{cv\blk}\;,
\label{eigeneqX}
\end{align}
with
\begin{align}
K^{\blQ}_{cv\blk,c'v'\blk'}=
W_{c\blk+\blQ\,v'\blk'\,v\blk\,c'\blk'+\blQ}-
v_{c\blk+\blQ\,v'\blk'\,c'\blk'+\blQ\,v\blk}.
\label{hsex}
\end{align}
Thus, a pure T-matrix self-energy (containing only 
$W$) cannot yield the
Lehmann intensity of the signal.  

\begin{figure}[t]
    \centering
\includegraphics[width=0.48\textwidth]{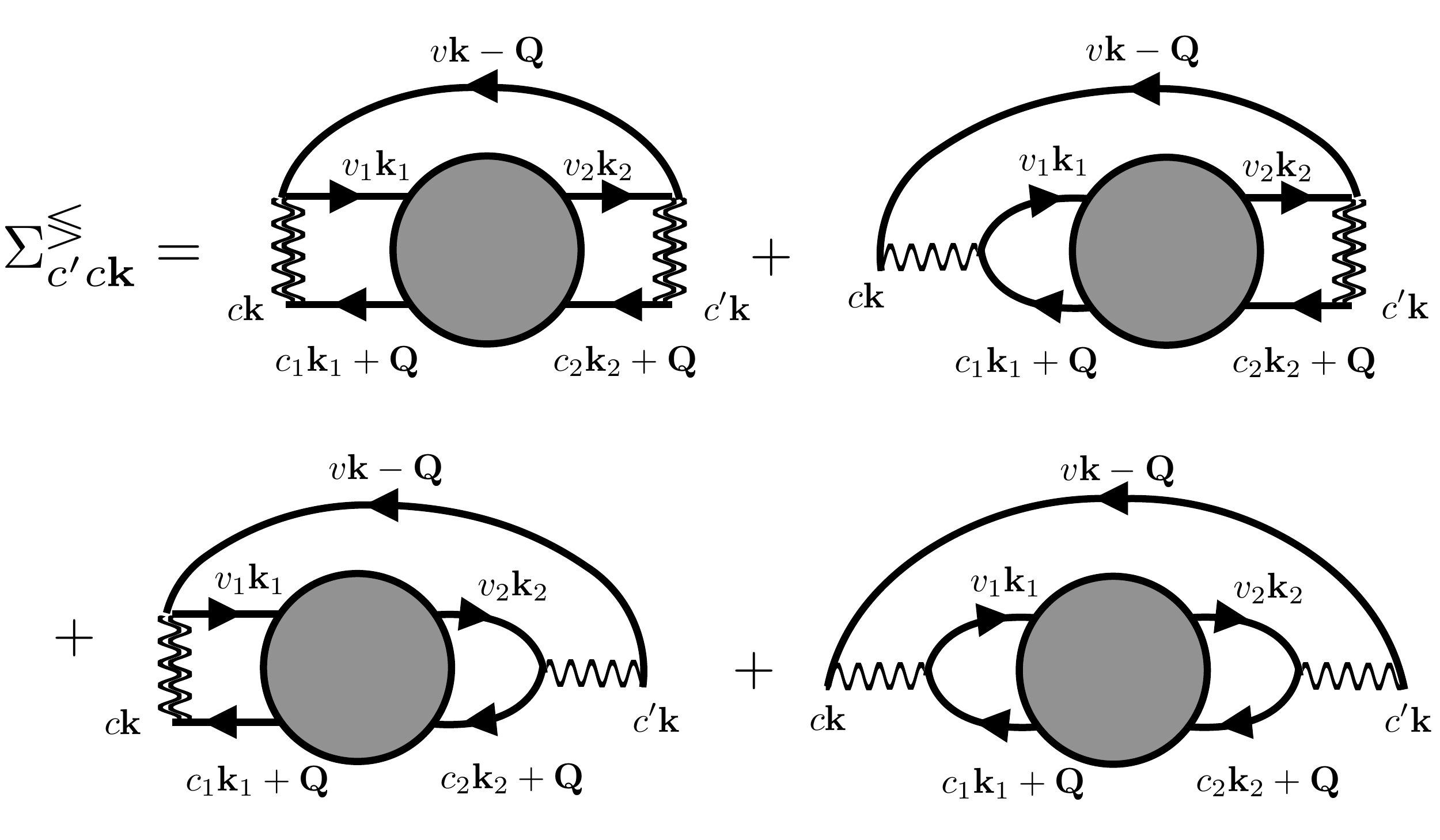}
\caption{TX self-energy for the Green's function in the incoherent 
regime. We represent Green's functions $G$ with solid 
lines, bare interactions $v$ with wiggly lines, screened interactions 
$W$ with wiggly lines and the xc function $L$ with 
a gray circle.}
\label{Tmatself}
\end{figure}

To reconcile many-body theory with the  Lehmann approach we consider the four self-energy 
diagrams of Fig.~\ref{Tmatself}  -- the gray 
circle is the exchange-correlation (xc) function $L$, the wiggly line 
is the bare interaction $v$, and the double wiggly line is the 
statically screened interaction $W$.
The first diagram in Fig.~\ref{Tmatself} is the T-matrix self-energy, 
and  we refer to the sum of all four diagrams as the T-matrix plus 
exchange (TX) self-energy. 
Notice that the zeroth order xc function is simply $GG$. Consequently, 
the lowest-order T-matrix diagram
has the same topology as the (second-Born) bubble diagram, 
except that {\em both} interaction lines are screened. 
This raises the issue of a possible double counting of screening 
diagrams, which we address below.

The diagrammatic equation of Fig.~\ref{Tmatself} is
an approximation for the {\em lesser} and {\em greater} self-energy, 
not for the time (or contour) ordered $\S$. Accordingly, 
the bubble diagram with two $W$ lines does not emerge from T-matrix, 
rather from GW. Indeed, 
$\S_{\rm GW}^{<}=iG^{<}W^{<}=iG^{<}W^{R}P^{<}W^{A}$ where 
$P^{<}=iG^{<}G^{>}$~\cite{svl-book_2025}.
Evaluating the screened interaction in 
the static approximation, $W^{R}=W^{A}=W$, we recover  
the lowest order (first)
diagram in Fig.~\ref{Tmatself}. Thus, there is no double counting 
issue. We also observe that 
the first diagram does not include 
the {\em full} GW self-energy, since the bottom Green's function
is summed over  
conduction bands only. 
The missing contributions are recovered by adding the 
fourth diagram in Fig.~\ref{Tmatself}. In fact, this diagram has the 
structure $Gv\chi v=G(W-v)$, $\chi$ being the density response 
function, and 
$G$ is summed over the valence bands. In summary,  the 
lesser and greater self-energies of Fig.~\ref{Tmatself} incorporate the full GW self-energy, 
but handle $W$ differently depending on the Green's function band indices. 
If $G$ has conduction indices then 
$W$ is statically screened, whereas for valence indices $W$ is 
dynamically screened.

Taking into account the definition of the HSEX kernel in 
Eq.~(\ref{hsex}), the TX self-energy of Fig.~\ref{Tmatself} 
can be written as
\begin{align}
\S^{\lessgtr}_{cc'\blk}(t,t')&=-i^{2}
\sum_{\substack{c_{1}c_{2}v_{1}v_{2}v\\\blk_{1}\blk_{2}\blQ}}
K^{\blQ}_{cv\blk-\blQ,c_{1}v_{1}\blk_{1}}
L^{\blQ\lessgtr}_{c_{1}v_{1}\blk_{1},c_{2}v_{2}\blk_{2}}(t,t')
\nn\\ &\times
K^{\blQ\ast}_{c'v\blk-\blQ,c_{2}v_{2}\blk_{2}}G^{\lessgtr}_{v\blk-\blQ}(t,t'),
\label{sigma<>}
\end{align}
from which we can extract the retarded self-energy according to
\begin{align}
\S^{R}_{cc'\blk}(\t,\w)&=i\int\frac{d\w'}{2\p}
\frac{\S^{>}_{cc'\blk}(\t,\w)-\S^{<}_{cc'\blk}(\t,\w)}{\w-\w'+i\eta}-i\eta.
\label{sigmaRTX}
\end{align}
Expanding the xc function $L$ in the excitonic 
basis~\cite{perfetto_first-principles_2016,stefanucci_excitonic_2025}, 
and approximating the 
Green's function at the quasiparticle level,
we obtain Eq.~(\ref{sigmaR}) with $\s$ evaluated in $N_{\l\blQ}=N^{\rm 
inc}_{\l\blQ}$ -- we refer the reader to 
Appendix~\ref{txapp} for the derivation.
This result agrees with that of a pure T-matrix 
calculation~\cite{steinhoff_exciton_2017} only if the exciton
energies and wavefunctions 
remain unchanged upon setting 
the bare $v$ to zero
in Eq.~(\ref{hsex}), which is generally not the case.

Equation~(\ref{sigma<>}) is only used to
extract $\S^{R}$. 
The lesser self-energy in Eq.~(\ref{G<}) is derived from 
 $\s$ according to Eq.~(\ref{sigma<}), ensuring consistency with the 
self-energy in the coherent regime, see Appendix~\ref{hsexmultbandapp}. 
As we see below [Eq.~(\ref{G<i2})], such consistency 
is essential not only for recovering the results of the Lehmann 
approach but also for extending the theory beyond the low-density 
limit.

{\em Low density.--}
In the low-density limit  $N_{\l\blQ}$ is infinitesimal, and
$E_{\l\blQ}(\t)$ and $\e_{\m\blk}(\t)$ 
are well approximated by their equilibrium values.
To lowest order in the 
excitation density, Eq.~(\ref{G<}) yields
\begin{align}
G^{<}_{cc'\blk}&(\t,\w)=\lim_{\eta\to 0}\frac{1}{\w-\e^{\rm eq}_{c\blk}+i\eta}
\S^{<}_{cc'\blk}(\t,\w)\frac{1}{\w-\e^{\rm eq}_{c'\blk}-i\eta}
\nn\\
=&2\p i\sum_{\l\blQ v}N_{\l\blQ}(\t)\;A^{\l\blQ}_{cv\blk-\blQ}
A^{\l\blQ\ast}_{c'v\blk-\blQ}\d(\w-E^{\rm eq}_{\l\blQ}-\e^{\rm eq}_{v\blk-\blQ}),
\label{G<tmatfinld}
\end{align}
where we use Eq.~(\ref{contrdef}).
This result agrees with the Lehmann approach both in the coherent 
($N_{\l\blQ}=\d_{\blQ\bz}|\r_{\l}|^{2}$) and 
incoherent ($N_{\l\blQ}=N_{\l\blQ}^{\rm inc}$) regimes, see 
Appendix~\ref{lehmannapp}.
It also agrees with the 
removal part of the spectral function  in the incoherent regime presented in 
Ref.~\cite{steinhoff_exciton_2017}.

\begin{figure*}[tbp]
    \centering
\includegraphics[width=0.98\textwidth]{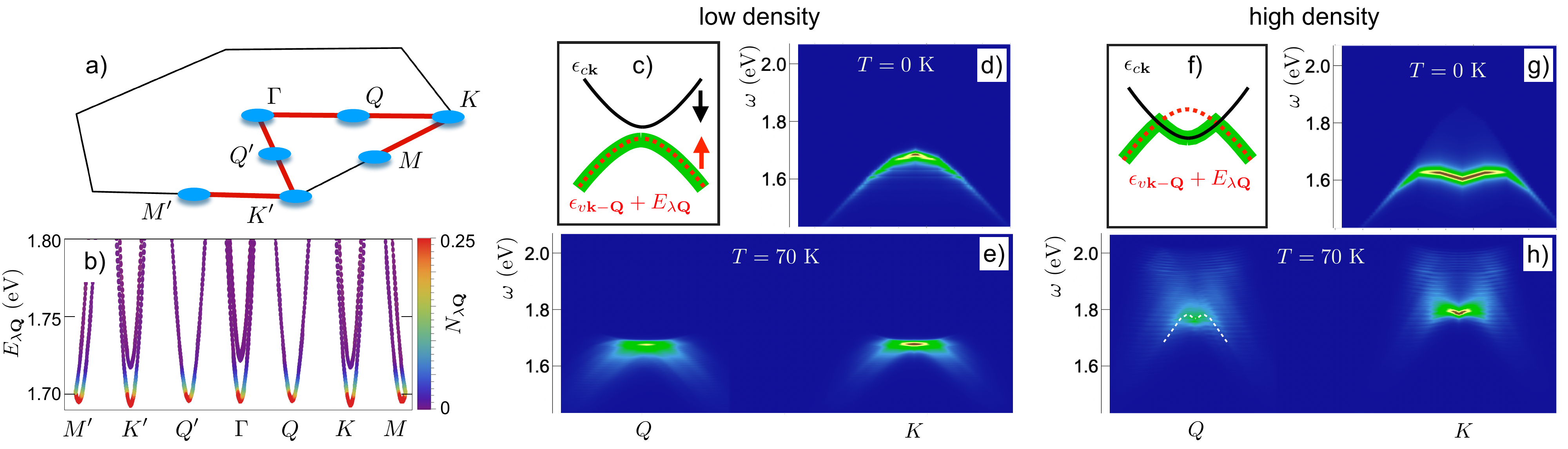}
\caption{(a) First Brillouin zone and location of a few high symmetry 
points. (b) Exciton band structure of the equilibrium  WSe$_{2}$ ML 
along the red path illustrated in panel~(a) -- curves are colored 
according to the thermal occupation of excitons at temperature 
$T=70$~K and excitation density $n_{c}=10^{13}$~cm$^{-2}$. (c,f)
Illustration of conduction band and replica of valence band at low 
[panel~(c)] and moderate [panel~(f)]
excitation densities. (d,e,g,h) TR-ARPES spectrum (in arbitrary 
units) for low [panels~(d,e)] and high [panels~(g,h)] excitation 
densities at different temperatures: 
$n_{c}=10^{11}$~cm$^{-2}$ and $T=0$~K [panel~(d)];  
$n_{c}=10^{11}$~cm$^{-2}$ and  $T=70$~K 
[panel~(e)]; $n_{c}=4\times 10^{12}$~cm$^{-2}$ and 
$T=0$~K [panel~(g)];  $n_{c}=10^{13}$~cm$^{-2}$ and  $T=70$~K [panel~(h)].
} 
\label{spectrafig}
\end{figure*}

Inserting Eq.~(\ref{G<tmatfinld}) in the photocurrent formula 
Eq.~(\ref{tdph2}) we find
\begin{align}
I_{\blk}(\t,\e)&\propto |a|^{2}\sum_{\l\blQ}N_{\l\blQ}(\t)\sum_{v}
\Big|\sum_{c}D_{c\blk}(\e)A^{\l\blQ}_{cv\blk-\blQ}\Big|^{2}
\nn\\
&\times 2\p\d(\e^{\rm eq}_{v\blk-\blQ}+E^{\rm eq}_{\l\blQ}-\e+\w_{0}).
\label{tdphincoh1}
\end{align}	
This formula extends the theory of 
Refs.~\cite{perfetto_first-principles_2016,rustagi_photoemission_2018} 
to multiple bands, and corrects the ansatz in 
Ref.~\cite{chen_first-principles_2022} by placing the sum over 
conduction bands inside the square modulus.
According to Eq.~(\ref{tdphincoh1}), a $\l\blQ$ exciton generates a 
replica of the valence bands shifted in energy by $E^{\rm eq}_{\l\blQ}$ 
and in momentum by $\blQ$, see Fig.~\ref{spectrafig}(c).
It is worth observing that if the lowest-energy exciton, say $\l_{0}$, 
is bright, and if we pump resonantly with it, then 
$\r_{\l}(\t)\propto \d_{\l\l_{0}}$ (resonance condition).
As phonons do not scatter the exciton into higher energy states, the 
coherent $\l_{0}$-exciton can only transition into an incoherent  
$\l_{0}$-exciton of vanishing momentum. Considering the conservation of the total number of 
excitons~\cite{stefanucci_excitonic_2025}, we infer that 
$N_{\l\blQ}(\t)\propto \d_{\l\l_{0}}\d_{\blQ\bz}$ is  independent of $\t$. 
Thus, in this scenario, the TR-ARPES signal is identical 
in the coherent and incoherent 
regimes~\cite{notepolaronic}.
Revealing the coherence of the system with TR-ARPES 
requires a 
probe field with a duration comparable to the exciton 
period~\cite{perfetto_time-resolved_2020}.     

An important feature of Eq.~(\ref{G<tmatfinld}) is that the total 
number of carriers in the conduction band is given by 
\begin{align}
N_{c}(\t)=-i\int\frac{d\w}{2\p}\sum_{c\blk}G^{<}_{cc\blk}(\t,\w)=
\sum_{\l\blQ}N_{\l\blQ}(\t),
\label{nc=nx}
\end{align}
where the normalization of the exciton wavefunctions 
$\sum_{cv\blk}|A^{\l\blQ}_{cv\blk}|^{2}=1$ has been used.
In other words, in the low density limit $N_{c}$ is the same as the 
total number of excitons, as it should be. 
We remark that this property would not be satisfied if we  had
included only the T-matrix diagram in the self-energy. 

{\em Beyond low density.--}
The formula Eq.~(\ref{G<}) for $G^{<}$  simplifies if the wavefunction 
of the occupied exciton states
have a predominant contribution on a single conduction 
band and valence band, say $c_{0}$ and $v_{0}$. In this case, 
$K^{\l\blQ}_{cv\blk}=\d_{cc_{0}}\d_{vv_{0}}K^{\l\blQ}_{c_{0}v_{0}\blk}$, 
and the quantity $\s$ in Eq.~(\ref{sigmasmall}) has one single 
nonvanishing entry for $c=c'=c_{0}$. This implies that also the 
self-energy is nonvanishing only for $c=c'=c_{0}$. Defining 
$\D_{\blk}^{\l\blQ}(\t)\equiv 
\sqrt{N_{\l\blQ}(\t)}K_{c_{0}v_{0}\blk}^{\l\blQ}$, and omitting the 
depence of the various quantities on $\t$, we have 
\begin{align}
G^{R}_{c_{0}c_{0}\blk}(\w)=\frac{1}{\w-\e_{c_{0}\blk}-
\sum_{\l\blQ}
\frac{|\D_{\blk}^{\l\blQ}|^{2}}{\w-E_{\l\blQ}-\e_{v_{0}\blk-\blQ}+i\eta}+i\eta}.
\label{twoband}
\end{align}
To gain insight into this problem,  we assume that 
only a discrete number of excitons $\l_{j}\blQ_{j}$ are populated, 
and that $ \D_{\blk}^{j}\equiv \D_{\blk}^{\l_{j}\blQ_{j}}$ are nonvanishing in small and 
nonoverlapping regions $\callR^{j}$ 
of the first Brillouin zone. Then, for  $\blk\in \callR^{j}$ only  
$\l_{j}\blQ_{j}$ contributes to the sum in the denominator of 
Eq.~(\ref{twoband}). This allows us to calculate 
$G^{<}_{c_{0}c_{0}\blk}$ analytically, see Appendix~\ref{hsextwobandapp} for details, and 
find
\begin{align}
G^{<}_{c_{0}c_{0}\blk}(\w)\simeq 2\p i  \sum_{j}  
\frac{E_{\l_{j}\blQ_{j}}+\e_{v_{0}\blk-\blQ_{j}}-E_{\blk,j}^{-}}
{E_{\blk,j}^{+}-E_{\blk,j}^{-}}\d(\w-E_{\blk,j}^{-}),
\label{G<i2}
\end{align}
where 
\begin{align}
E_{\blk,j}^{\pm}&=\frac{1}{2}
\Big[\e_{c_{0}\blk}+E_{\l_{j}\blQ_{j}}+\e_{v_{0}\blk-\blQ_{j}}
\nn\\
&\pm\sqrt{(\e_{c_{0}\blk}-E_{\l_{j}\blQ_{j}}
-\e_{v_{0}\blk-\blQ_{j}})^{2}+4|\D_{\blk}^{j}|^{2}}\Big].
\label{epm}
\end{align}

To lowest order in $|\D^{j}_{\blk}|$ 
(low density limit) we can approximate the argument of the 
Dirac-delta as $E_{\blk,j}^{-}\simeq
E^{\rm eq}_{\l_{j}\blQ_{j}}+\e^{\rm eq}_{v_{0}\blk-\blQ_{j}}$,
and the prefactor of the Dirac-delta as 
$|\D_{\blk}^{j}|^{2}/(\e^{\rm eq}_{c_{0}\blk}-E^{\rm eq}_{\l_{j}\blQ_{j}}
-\e^{\rm eq}_{v_{0}\blk-\blQ_{j}})^{2}=N_{\l_{j}\blQ_{j}}|A^{\l_{j}\blQ_{j}}_{c_{0}v_{0}\blk}|^{2}$; 
the resulting $G^{<}$ correctly agrees with 
Eq.~(\ref{G<tmatfinld}).
Increasing the excitation density, the nonequilibrium 
energy $E_{\l_{j}\blQ_{j}}(\t)+\e_{v_{0}\blk-\blQ_{j}}(\t)$ 
might overcome $\e_{c_{0}\blk}(\t)$, leading to a  
hybridization (proportional to $|\D^{j}_{\blk}|$) between the 
conduction band and the replica of the valence band, see 
Fig.~\ref{spectrafig}(f). 
This scenario has been thoroughly investigated in the 
coherent 
regime~\cite{perfetto_pump-driven_2019,perfetto_time-resolved_2020,chan_giant_2023} -- hence for $\blQ=\bz$ --
and is associated with a BEC-BCS crossover of the coherent exciton 
fluid~\cite{ostreich_non-stationary_1993,hannewald_excitonic_2000}.
Our result in Eq.~(\ref{G<i2}) 
indicates that the same phenomenon can also occur in the 
incoherent regime. In particular, the excitonic sidebands induced by 
a sizable population of incoherent 
excitons of momentum $\blQ$ might resemble the conduction 
bands hybridized with a $\blQ$-shifted replica of the valence bands.

{\em TR-ARPES in WSe$_{2}$ monolayer.--} We consider a WSe$_{2}$ 
monolayer (ML) at low ($n_{c}\sim 10^{11}$~cm$^{-2}$) and moderate 
($n_{c}\sim 10^{13}$~cm$^{-2}$) excitation densities $n_{c}=N_{c}/A$, 
where $A$ is the area of the sample. The WSe$_{2}$ ML is a direct gap semiconductor, 
with a 
calculated bandgap at the $K$ and $K'$ points [Fig.~\ref{spectrafig}(a)] 
estimated to be $\sim 2$~eV~\cite{lin_narrow-band_2021}. 
For sufficiently long times after pumping, the exciton populations follow a  
Bose distribution 
$N_{\l\blQ}=[e^{(E_{\l\blQ}-\m_{X})/T}-1]^{-1}$, where $T$ is the 
temperature and $\m_{\rm X}$ is self-consistently calculated from 
Eq.~(\ref{nc=nx}).
In Fig.~\ref{spectrafig}(b) we show the equilibrium
exciton band structure, and how excitons are distributed 
at temperature $T=70$~K; numerical details are reported in 
Appendix~\ref{numapp}. 
The low-energy landascape is characterized by several (almost) degenerate 
excitons~\cite{deilmann_finite-momentum_2019,madeo_directly_2020}, 
the lowest excitons being at $K$ and $K'$.   

At zero temperature, only the degenerate excitons with momenta $\blQ$ 
at $K$ and $K'$ are populated. Along the path $\G-K-\G$, and 
for low values of $n_{c}$, the spectrum features a single sideband 
for momenta $\blk$ near $K$ [Fig.~\ref{spectrafig}(d)]. 
This is the replica of $\e_{v\blk-\blQ}$ with $\blQ$ 
at the inequivalent $K$ point, which is the same as 
$\e_{v\blk}$ with $\blk$ around $K'$. The replica is shifted in 
energy by $\sim E_{\l\blQ}^{\rm eq}$, and the
intensity of the signal is proportional to $|A^{\l\blQ}_{cv\blk-\blQ}|^{2}$, in 
agreement with Eq.~(\ref{G<tmatfinld}). Increasing $n_{c}$,
the exciton energies undergo a blue shift while the  
bandgap shrinks [Figs.~\ref{spectrafig}(c) and (f)]. 
The replica of the valence band hybridize with the 
conduction band, and the excitonic sideband acquires the shape of a 
mexican hat [well described by Eq.~(\ref{epm})], as shown in 
Fig.~\ref{spectrafig}(g). 

At temperatures $T=70$~K, exciton states with momenta near other high 
symmetry points ($M$, 
$Q$ and $\G$) also become 
populated, see  Fig.~\ref{spectrafig}(b). 
Along the path $\G-K-\G$, and 
for low values of $n_{c}$, a second 
replica of the valence band appears for momenta $\blk$ near
$Q$ [Fig.~\ref{spectrafig}(e)]. This emerges from 
excitons with $-\blQ$ around $Q$, leading to a replica of $\e_{v\blk}$ 
with $\blk$ around $K$. As the exciton energies are almost 
degenerate, the two replicas in 
Fig.~\ref{spectrafig}(e) are located in the same spectral range.
Compared to Fig.~\ref{spectrafig}(d), the signal appears 
blurred -- this is due to the presence of exciton states with momenta close to, 
but not exactly at, the high symmetry points. The additional excitonic 
sideband around the $Q$ point persists at higher excitation 
densities, but its shape turns into a mexican hat, see 
Fig.~\ref{spectrafig}(h), again in agreement with Eq.~(\ref{epm}).

{\em Conclusions.--} We presented a first-principles approach to 
excitons in TR-ARPES, bridging the coherent and incoherent regimes 
through a unified formula that depends solely on the total number of 
excitons at the pump-probe delay $\t$.
Applicable across a wide 
range of temperatures and  
excitation densities, our formula provides a powerful  
tool for understanding and engineering nonequilibrium band structures 
of excitonic materials. 
Band hybridization with dark excitons enables 
controlled shifts of the valence bands in both energy and momentum, 
and to reshape them into a mexican-hat dispersion.
In particular, 
the mechanism of 
Ref.~\cite{foster_quench-induced_2014,perfetto_floquet_2020} 
naturally extends to  the incoherent regime, enabling 
control over the topological character of excited materials 
through exciton wavefunction symmetries.
This points to a 
promising avenue for realizing nonequilibrium
topological phases with 
lifetimes reaching the picosecond timescale.

\acknowledgments

We acknowledge insightful discussions with Keshav Dani and Felipe da 
Jornada.
We acknowledge funding from Ministero Universit\`a e 
Ricerca PRIN under grant agreement No. 2022WZ8LME, 
from INFN through project TIME2QUEST, 
from European Research Council MSCA-ITN TIMES under grant agreement 101118915, 
and from Tor Vergata University through project TESLA.

\appendix
\begin{widetext}

\section{Finite excitation-density corrections}
\label{findensapp}

We here describe a practical method to estimate the finite-density 
corrections to the quasi-particle and exciton energies from the sole 
knowledge of the exciton populations $N_{\l\blQ}$.

Let $\r_{\m\m'\blk}(t)=-iG_{\m\m'\blk}^{<}(t,t)$ be the one-electron 
density matrix at time $t$. We define 
$\r^{\rm eq}_{\m\m'\blk}=\d_{\m\m'}f^{\rm eq}_{\m\blk}$ as the
equilibrium one-electron  density matrix, with equilibrium occupations $f^{\rm 
eq}_{v\blk}=1$ and $f^{\rm eq}_{c\blk}=0$. The change of the density 
matrix  caused by an external pump pulse is then
\begin{align}
\D\r_{\m\m'\blk}(t)=\r_{\m\m'\blk}(t)-\r^{\rm eq}_{\m\m'\blk}.
\end{align}
This change induces the following change in the HSEX potential
\begin{subequations}
\begin{align}
V_{cc'\blk}(t)&=\sum_{vv'\blk'}\big(
v_{c\blk\, v\blk'\, v'\blk' \,c'\blk}-W_{c\blk\, v\blk'\,  c'\blk\, v'\blk'}
\big)\D\r_{v'v\blk'}(t)
+\sum_{c_{1}c_{2}\blk'}\big(
v_{c\blk\, c_{1}\blk'\, c_{2}\blk' \,c'\blk}-W_{c\blk\, c_{1}\blk'\,  
c'\blk\, c_{2}\blk'}
\big)\D\r_{c_{2}c_{1}\blk'}(t),
\\
V_{vv'\blk}(t)&=\sum_{cc'\blk'}\big(
v_{v\blk\, c\blk'\, c'\blk'\, v'\blk}-W_{v\blk\, c\blk'\,  v'\blk\, c'\blk'}
\big)\D\r_{c'c\blk'}(t)
+\!\sum_{v_{1}v_{2}\blk'}\!\big(
v_{v\blk\, v_{1}\blk'\, v_{2}\blk'\, v'\blk}-W_{v\blk\, v_{1}\blk'\,  
v'\blk\, v_{2}\blk'}
\big)\D\r_{v_{2}v_{1}\blk'}(t),
\\
V_{cv\blk}(t)&=\sum_{v'c'\blk'}\big(
v_{c\blk\, v'\blk'\, c'\blk'\, v\blk}-W_{c\blk\, v'\blk'\,  v\blk\, c'\blk'}
\big)\D\r_{c'v'\blk'}(t)
=-\sum_{v'c'\blk'}K^{\bz}_{cv\blk,c'v'\blk'}\D\r_{c'v'\blk'}(t)
=V_{vc\blk}^{\ast}(t).
\label{hsexpotcv}
\end{align}
\label{hsexpot}
\end{subequations}
In the HSEX approximation the renormalization of the quasiparticle 
energies can be estimated as
\begin{align}
\e_{\m\blk}=\e^{\rm eq}_{\m\blk}+V_{\m\m\blk},
\label{rqpene}
\end{align}
where	
\begin{subequations}
\begin{align}
V_{cc\blk}&\simeq \sum_{v'\blk'}\big(
v_{c\blk\, v'\blk'\, v'\blk' \,c\blk}-W_{c\blk\, v'\blk'\,  c\blk\, v'\blk'}
\big)(f_{v'\blk'}-1)+\sum_{c'\blk'}\big(
v_{c\blk\, c'\blk'\, c'\blk' \,c\blk}-W_{c\blk\, c'\blk'\,  
c\blk\, c'\blk'}
\big)f_{c'\blk'},
\label{epscR}
\\
V_{vv\blk}&\simeq \sum_{c'\blk'}\big(
v_{v\blk\, c'\blk'\, c'\blk'\, v\blk}-W_{v\blk\, c'\blk'\,  v\blk\, c'\blk'}
\big)f_{c'\blk'}
+\sum_{v'\blk'}\big(
v_{v\blk\, v'\blk'\, v'\blk'\, v\blk}-W_{v\blk\, v'\blk'\,
v\blk\, v'\blk'}\big)(f_{v'\blk'}-1).
\label{epsvR}
\end{align}
\label{epscvR}
\end{subequations}
In Eqs.~(\ref{epscvR}), the off-diagonal contributions of the density 
matrix are discarded, i.e.,  
$\D\r_{cc'\blk}\simeq \d_{cc'}f_{c\blk}$ and 
$\D \r_{vv'\blk}\simeq \d_{vv'}(f_{v\blk}-1)$.
This is consistent with ignoring  
finite-density corrections to the 
quasiparticle wavefunctions. Thus, Eq.~(\ref{epscvR}) provides a 
formula for the renormalized quasi-particle 
energies in terms of the nonequilibrium occupations $f_{\m\blk}$.

Let us now come to the renormalization of the exciton energies.
At finite excitation density, the BSE for the 
exciton energies and wavefunctions reads
\begin{align}
\sum_{c'v'\blk'}H^{\blQ}_{cv\blk,c'v'\blk'}
\frac{A^{\l\blQ}_{c'v'\blk'}}
{\sqrt{f_{v'\blk'}-f_{c'\blk'+\blQ}}}=
E_{\l\blQ}\frac{A^{\l\blQ}_{cv\blk}}{\sqrt{f_{v\blk}-f_{c\blk+\blQ}}},
\label{bse1}
\end{align}
with  
\begin{align}
H^{\blQ}_{cv\blk,c'v'\blk'}&=\d_{cv\blk,c'v'\blk'}
(\e_{c\blk+\blQ}-\e_{v\blk})
-\sqrt{f_{v\blk}-f_{c\blk+\blQ}}\,K^{\blQ}_{cv\blk,c'v'\blk'}
\sqrt{f_{v'\blk'}-f_{c'\blk'+\blQ}}
\label{bse2}
\end{align}
the BSE Hamiltonian.
We rewrite Eq.~(\ref{bse2}) as
\begin{align}
H^{\blQ}_{cv\blk,c'v'\blk'}=H^{{\rm eq},\blQ}_{cv\blk,c'v'\blk'}+\D 
H^{\blQ}_{cv\blk,c'v'\blk'},
\end{align}
where $H^{{\rm eq},\blQ}$ is the equilibrium BSE Hamiltonian,
calculated with equilibrium occupations $f^{\rm eq}_{v\blk}=1$ and 
$f^{\rm eq}_{c\blk}=0$, and  
equilibrium band structure $\e_{\m\blk}=\e^{\rm eq}_{\m\blk}$.
To lowest order in $\D H^{\blQ}$ we have
\begin{align}
E_{\l\blQ}=E^{\rm eq}_{\l\blQ}+\sum_{cv\blk}\sum_{c'v'\blk'}
A^{\l\blQ\ast}_{cv\blk}\D 
H^{\blQ}_{cv\blk,c'v'\blk'}A^{\l\blQ}_{c'v'\blk'},
\label{rxene}
\end{align}
where $E^{\rm eq}_{\l\blQ}$ is the equilibrium exciton energy and 
$A^{\l\blQ}_{cv\blk}$ is the unperturbed exciton wavefunction.

We express $\D H^{\blQ}$ in terms of the nonequilibrium occupations 
$f_{\m\blk}$.
For the Pauli blocking factors we have
\begin{align}
\sqrt{f_{v\blk}-f_{c\blk+\blQ}}
\simeq 
1+\frac{1}{2}\big(f_{v\blk}-f_{c\blk+\blQ}-1\big).
\end{align}
We conclude that to lowest order in the excitation density
\begin{align}
\D H^{\blQ}_{cv\blk,c'v'\blk'}&=
\d_{cv\blk,c'v'\blk'}\Big[V_{cc\blk+\blQ}-V_{vv\blk}\Big]
-\frac{1}{2}\big(f_{v\blk}-f_{c\blk+\blQ}+
f_{v\blk'}-f_{c\blk'+\blQ}-2\big)
K^{\blQ}_{cv\blk,c'v'\blk'},
\label{deltabse}
\end{align}
where $V_{\m\m\blk}$ is given in Eq.~(\ref{epscvR}).

Equations~(\ref{rqpene}) and (\ref{rxene}) provide an approximation for the 
nonequilibrium quasi-particle energies and exciton energies in terms 
of the 
nonequilibrium occupations. To  express these occupations in terms of exciton 
populations we observe that for small excitation densities 
the contribution of the $\l\blQ$-exciton to  $f_{c\blk}$ is 
\begin{align}
\bra\l\blQ|\hat{d}^{\dag}_{c\blk}
\hat{d}_{c\blk}|\l\blQ\ket
&=\sum_{c_{1}v_{1}\blk_{1}}\sum_{c_{2}v_{2}\blk_{2}}
A^{\l\blQ\ast}_{c_{1}v_{1}\blk_{1}}
A^{\l\blQ}_{c_{2}v_{2}\blk_{2}}
\bra\Q_{g}|\hat{d}^{\dag}_{v_{1}\blk_{1}}
\hat{d}_{c_{1}\blk_{1}+\blQ}\hat{d}^{\dag}_{c\blk}
\hat{d}_{c\blk}\hat{d}^{\dag}_{c_{2}\blk_{2}+\blQ}
\hat{d}_{v_{2}\blk_{2}}|\Q_{g}\ket
=\sum_{v}|A^{\l\blQ}_{cv\blk-\blQ}|^{2}.
\end{align}
Therefore
\begin{align}
f_{c\blk}=\sum_{\l\blQ 
v}N_{\l\blQ}|A^{\l\blQ}_{cv\blk-\blQ}|^{2},
\label{fck}
\end{align}
and similarly
\begin{align}
f_{v\blk}=1-\sum_{\l\blQ 
c}N_{\l\blQ}|A^{\l\blQ}_{cv\blk}|^{2}.
\label{fvk}
\end{align}
Notice that the normalization of the exciton wavefunctions
guarantees that 
$\sum_{c\blk}f_{c\blk}=\sum_{\l\blQ}N_{\l\blQ}$.

In Fig.~\ref{finite-corrections} we compare the equilibrium band structure and exciton 
dispersions with those at moderate excitation density $n_{c}\simeq 
10^{13}$~cm$^{-2}$ for the monolayer WSe$_{2}$. Due to finite density corrections 
the replica of high-energy valence bands intersect the low-energy 
conduction bands. In this scenario, the exciton-induced hybridization 
causes the typical mexican hat shape in TR-ARPES. 

\begin{figure}[t]
    \centering
\includegraphics[width=0.7\textwidth]{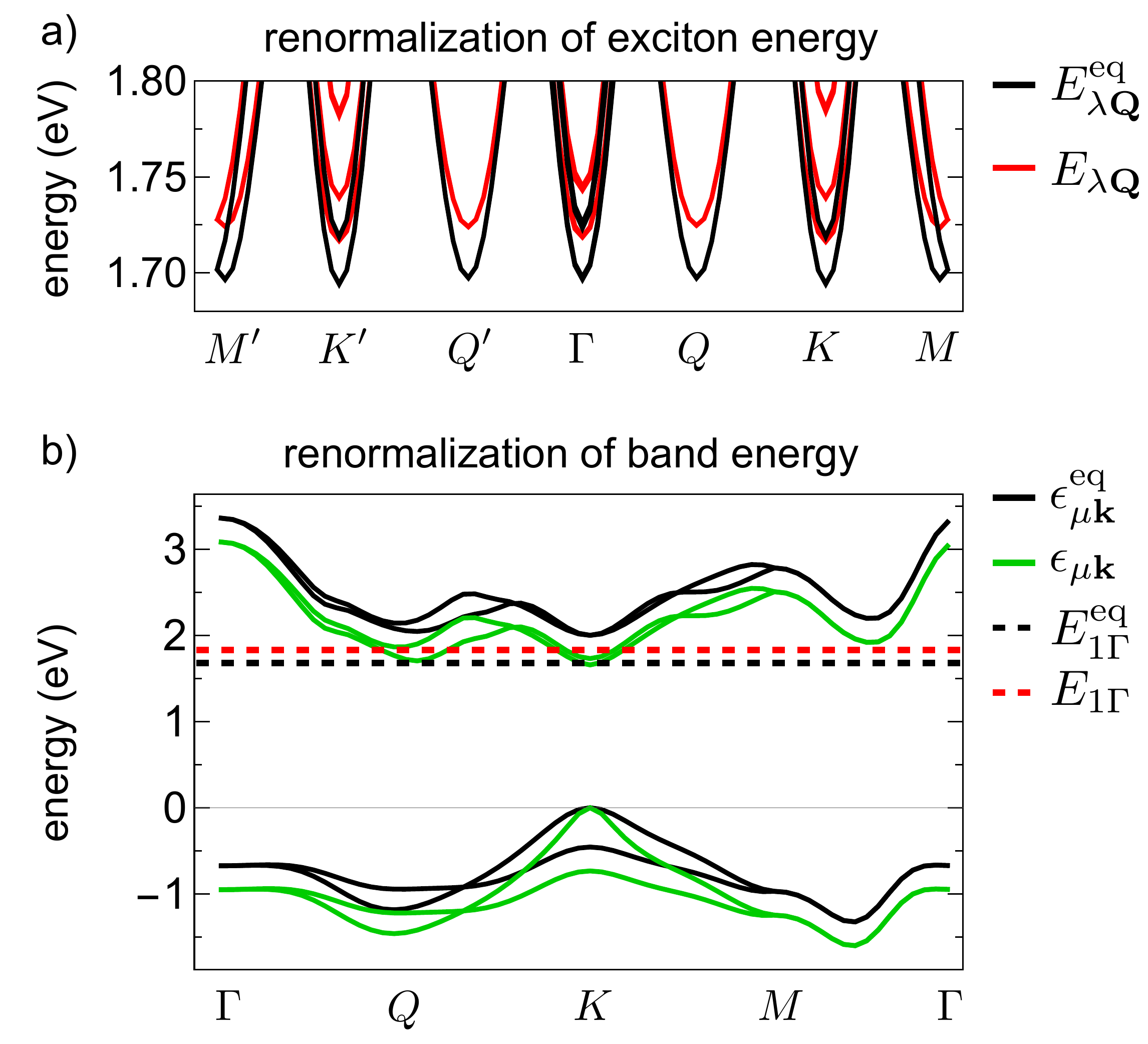}
\caption{Equilibrium and nonequilibrium [excitation density $n\simeq 
10^{13}$~cm$^{-2}$] exciton dispersions (a)
and quasiparticle energies (b). To highlight the band crossing between 
the conduction bands and the replica of the valence bands, dashed 
horizontal lines have been superimposed in correspondence of the 
energy of the lowest exciton with vanishing momentum.}
\label{finite-corrections}
\end{figure}

\section{Lehmann approach}
\label{lehmannapp}
The state of the system just {\em after} pumping ($t>0$)
can be written as 
\begin{align}
|\Q(t)\ket=|\Q_{g}\ket+\sum_{\l}\b_{\l}e^{-iE^{\rm eq}_{\l\bz}t}|\l\bz\ket.
\label{cohstate}
\end{align}
The coefficients $\b_{\l}$ are proportional to the excitonic 
polarizations since
\begin{align}
\r_{\l}(t)&=
\sum_{cv\blk}A^{\l\bz\ast}_{cv\blk}\,
\bra\Q(t)|\hat{d}^{\dag}_{v\blk}\hat{d}_{c\blk}|\Q(t)\ket
=\sum_{cv\blk}A^{\l\bz\ast}_{cv\blk}\,
\sum_{\l'}\b_{\l'}A^{\l'\bz}_{cv\blk}e^{-iE^{\rm eq}_{\l'\bz}t}
=\b_{\l}e^{-iE^{\rm eq}_{\l\bz}t}.
\label{rldef}
\end{align}
Thus, Eq.~(\ref{cohstate}) can also be written as 
\begin{align}
|\Q(t)\ket=|\Q_{g}\ket+\sum_{\l}\r_{\l}(t)|\l\bz\ket.
\label{cohstate2}
\end{align}
For a pure state the Green's function is given by
\begin{align}
G^{<}_{\m\m'\blk}(t,t')\equiv i\bra\Q(t')|\hat{d}^{\dag}_{\m'\blk}e^{i\hat{H}(t-t')}
\hat{d}_{\m\blk}|\Q(t)\ket,
\label{g<def}
\end{align}
where $\hat{H}$ is the many-body Hamiltonian of the unperturbed system.
We have
\begin{align}
\hat{d}_{c'\blk}|\Q(t)\ket=
\sum_{\l}\b_{\l} e^{-iE^{\rm eq}_{\l\bz}t}\sum_{v}
A^{\l\bz}_{c'v\blk}\hat{d}_{v\blk}|\Q_{g}\ket.
\end{align}
Taking into account that $\hat{H}\hat{d}_{v\blk}|\Q_{g}\ket
=-\e^{\rm eq}_{v\blk}\hat{d}_{v\blk}|\Q_{g}\ket$,
we find for the Green's function in Eq.~(\ref{g<def})
\begin{align}
G^{<}_{cc'\blk}(t,t')=i\sum_{\l\l'v}\b_{\l}^{\ast}\b_{\l'}\;
A^{\l\bz\ast}_{c'v\blk}A^{\l'\bz}_{cv\blk}\;
e^{iE^{\rm eq}_{\l\bz}t'}e^{-iE^{\rm eq}_{\l'\bz}t}
e^{-i\e^{\rm eq}_{v\blk}(t-t')}.
\label{gfcohleh}
\end{align}
This Green's function depends on $t$ and $t'$ separately, due to the 
interference of different exciton 
states~\cite{rustagi_coherent_2019,zhang_long-lived_2025}. The Fourier 
transform with respect to $t-t'$ is dominated by the diagonal 
contributions and reads
\begin{align}
G^{<}_{cc'\blk}(\w)=2\p i\sum_{\l v}|\r_{\l}|^{2}\;
A^{\l\bz\ast}_{c'v\blk}A^{\l\bz}_{cv\blk}\;
\d(\w-E^{\rm eq}_{\l\bz}-\e^{\rm eq}_{v\blk}).
\label{gfcohlehw}
\end{align}

For low excitation densities, the coherence between $|\Q_{g}\ket$ and the exciton states
$|\l\bz\ket$ is suppressed by electron-phonon scatterings on a 
timescale that varies from a few tens to a few hundreds of 
femtoseconds (depending on temperature and excitation 
densities)~\cite{moody_intrinsic_2015,selig_exciton_2016,jakubczyk_impact_2018}. In the incoherent regime the 
polarization $\r_{\l}$ vanishes, and the 
electronic subsystem is described by an ensemble of bright and dark 
incoherent excitons. From the many-body density matrix of the ensemble 
$\hat{\r}=|\Q_{g}\ket\bra\Q_{g}|+\sum_{\l\blQ}N^{\rm 
inc}_{\l\blQ}|\l\blQ\ket\bra\l\blQ|$, with $N^{\rm inc}_{\l\blQ}$  
the number of incoherent excitons,  we can calculate  
the Green's function  according to
\begin{align}
G^{<}_{cc'\blk}(t,t')=i\Tr\big[\hat{\r}\,e^{i\hat{H}t'}\hat{d}^{\dag}_{\m'\blk}e^{i\hat{H}(t-t')}
\hat{d}_{\m\blk}e^{-i\hat{H}t}\big]=
\sum_{\l\blQ}N^{\rm 
inc}_{\l\blQ}G^{\l\blQ<}_{cc'\blk}(t,t'),
\label{g<ens}
\end{align}
where
\begin{align}
G^{\l\blQ<}_{cc'\blk}(t,t')&=i\bra\l\blQ(t')|
\hat{d}^{\dag}_{c'\blk}e^{i\hat{H}(t-t')}\hat{d}_{c\blk}|\l\blQ(t)\ket
\nn\\
&=ie^{-iE^{\rm eq}_{\l\blQ}(t-t')}
\bra\l\blQ|\hat{d}^{\dag}_{c'\blk}e^{i\hat{H}(t-t')}\hat{d}_{c\blk}|\l\blQ\ket
\nn\\
&=ie^{-iE^{\rm eq}_{\l\blQ}(t-t')}\sum_{v}
A^{\l\blQ\ast}_{c'v\blk-\blQ}A^{\l\blQ}_{cv\blk-\blQ}
e^{-i\e^{\rm eq}_{v\blk-\blQ}(t-t')}.
\label{G<isht}
\end{align}
In the incoherent regime the Green's function depends exclusively on 
the time difference, and its Fourier transform reads
\begin{align}
G^{<}_{cc'\blk}(\w)=2\p i\sum_{\l\blQ v}N^{\rm inc}_{\l\blQ}
A^{\l\blQ\ast}_{c'v\blk-\blQ}A^{\l\blQ}_{cv\blk-\blQ}\;
\d(\w-E^{\rm eq}_{\l\blQ}-\e^{\rm eq}_{v\blk-\blQ}).
\label{G<ishtw}
\end{align}
Notice that if the population of incoherent excitons is given by 
$N^{\rm inc}_{\l\blQ}=\d_{\blQ,\bz}|\r_{\l}|^{2}$, then the Green's 
function -- and, consequently, the TR-ARPES spectrum -- becomes 
identical for both the coherent and incoherent regimes. 
This can be verified by comparing Eq.~(\ref{G<ishtw}) with the expression for the 
Green's function in the coherent regime,  
i.e., Eq.~(\ref{gfcohlehw}). Equations~(\ref{gfcohlehw}) 
and~(\ref{G<ishtw}), derived from the Lehmann approach, 
agree with the Green's function in the many-body diagrammatic 
approximation HSEX+TX, see Eq.~(\ref{G<tmatfinld}) in the main text.

\section{HSEX self-energy for two bands}
\label{hsextwobandapp}

We consider a system with only one active conduction and valence 
bands. The HSEX lesser and greater Green's functions for 
resonant (not necessarily weak) pumping with the $\l$ 
exciton have been derived in Ref.~\cite{perfetto_pump-driven_2019}, and 
for sufficiently low temperatures they have a simple analytic form 
\begin{subequations}
\begin{align}
G^{<}_{cc\blk}(\w)&=2\p i
\frac{E_{\l\bz}+\e_{v\blk}-E^{-}_{\blk}}
{E^{+}_{\blk}-E^{-}_{\blk}} 
\d(\w-E^{-}_{\blk}),
\label{G<ccprm}
\\
G^{>}_{cc\blk}(\w)&=2\p i \frac{E_{\l\bz}+\e_{v\blk}-E^{+}_{\blk}}
{E^{+}_{\blk}-E^{-}_{\blk}} 
\d(\w-E^{+}_{\blk}),
\label{G>ccprm}
\end{align}
\label{Gccprm}
\end{subequations}
where
\begin{subequations}
\begin{align}
E^{\pm}_{\blk}&\equiv \frac{1}{2}\Big[
\e_{c\blk}+E_{\l\bz}+\e_{v\blk}
\pm S_{\blk}\Big],
\label{Eplusminus}
\\
S_{\blk}&\equiv\sqrt{(\e_{c\blk}-E_{\l\bz}-\e_{v\blk})^{2}
+4|\D_{\blk}|^{2}},
\\
\D_{\blk}&\equiv K^{\l\bz}_{cv\blk}\r_{\l}.
\end{align}
\end{subequations}
Setting $z=\w+i\eta$, the retarded Green's function is then given by
\begin{align}
G^{R}_{cc\blk}(\w)&=i\int\frac{d\w'}{2\p}	
\frac{G^{>}_{cc\blk}(\w')-G^{<}_{cc\blk}(\w')}{z-\w'}
=\frac{z-E_{\l\bz}-\e_{v\blk}}{
(z-E^{+}_{\blk})(z-E^{-}_{\blk})}
=\frac{1}{z-\e_{c\blk}-\frac{|\D_{\blk}|^{2}}{z-E_{\l\bz}-\e_{v\blk}}},
\label{GRanalyt2}
\end{align}
from which we can read the retarded HSEX self-energy
\begin{align}
\S^{R}_{cc\blk}(\w)=\frac{|\D_{\blk}|^{2}}{z-E_{\l\bz}-\e_{v\blk}}-i\eta.
\label{twobself}
\end{align}

To obtain the lesser HSEX self-energy we observe that 
the Green's function in Eq.~(\ref{G<ccprm}) can also be 
written as
\begin{align}
G^{<}_{cc\blk}(\w)=G^{R}_{cc\blk}(\w)\S^{<}_{cc\blk}(\w)G^{A}_{cc\blk}(\w).
\label{G<ccguess}
\end{align}
Let us show that this relation is satisfied for
\begin{align}
\S^{<}_{cc\blk}(\w)=-f(\w-E_{\l\bz}-\e_{v\blk})
\big[\S^{R}_{cc\blk}(\w)-\S^{A}_{cc\blk}(\w)\big],
\label{twobself2}
\end{align}
with $f(\w)=1/[e^{\w/T}+1]$ the Fermi function. 
We have 
\begin{align}
G^{R}_{cc\blk}(\w)\big[\S^{R}_{cc\blk}(\w)-\S^{A}_{cc\blk}(\w)\big]
G^{A}_{cc\blk}(\w)=G^{R}_{cc\blk}(\w)-G^{A}_{cc\blk}(\w)=
G^{>}_{cc\blk}(\w)-G^{<}_{cc\blk}(\w).
\end{align}
Therefore
\begin{align}
G^{R}_{cc\blk}(\w)\S^{<}_{cc\blk}(\w)G^{A}_{cc\blk}(\w)=
-2\p i f(\w-E_{\l\bz}-\e_{v\blk})
\Big[\frac{E_{\l\bz}+\e_{v\blk}-E^{+}_{\blk}}
{E^{+}_{\blk}-E^{-}_{\blk}} 
\d(\w-E^{+}_{\blk})-\frac{E_{\l\bz}+\e_{v\blk}-E^{-}_{\blk}}
{E^{+}_{\blk}-E^{-}_{\blk}} 
\d(\w-E^{-}_{\blk})\Big].
\label{gccguess2}
\end{align}
Next we observe that Eq.~(\ref{Eplusminus}) implies 
\begin{align}
E^{\pm}_{\blk}-E_{\l\bz}-\e_{v\blk}&=
\frac{1}{2}\Big[
\e_{c\blk}+E_{\l\bz}+\e_{v\blk}
\pm S_{\blk}\Big]-E_{\l\bz}-\e_{v\blk}
\nn\\
&=\frac{1}{2}\Big[
\e_{c\blk}-E_{\l\bz}-\e_{v\blk}
\pm \sqrt{(\e_{c\blk}-E_{\l\bz}-\e_{v\blk})^{2}
+4\D_{\blk}^{2}}\Big]\gtrless 0
\end{align}	
We conclude the second term in Eq.~(\ref{gccguess2}) dominates, and 
we recover the analytic result in Eq.~(\ref{G<ccprm}).

\section{HSEX self-energy for multiple bands}
\label{hsexmultbandapp}

In the general case, the Green's function in the HSEX approximation 
can be obtained from the 
equation of motion on the Keldysh contour~\cite{svl-book_2025}
\begin{align}
\Big[i\frac{d}{dz}-\e^{\rm eq}_{\m\blk}\Big]G_{\m\m'\blk}(z,z')
-\sum_{\n}V_{\m\n\blk}(t)G_{\n\m'\blk}(z,z')=\d(z,z'),
\label{eomGRmatk}
\end{align}
where the HSEX potential is given in Eqs.~(\ref{hsexpot}).
We look for an approximate solution of Eq.~(\ref{eomGRmatk}) in 
the conduction subspace, which is the subspace we are interested in.
We discard
all matrix elements $V_{cc'}$ with $c\neq c'$ 
and $V_{vv'}$ with $v\neq v'$. This is equivalent to discard 
finite-density corrections to the 
quasiparticle wavefunctions. Although these corrections do not pose a conceptual 
problem, they considerably complicate the practical implementation of 
the theory. From Eq.~(\ref{eomGRmatk}) we  have
\begin{subequations}
\begin{align}
\Big[i\frac{d}{dz}-\e_{c\blk}\Big]G_{cc'\blk}(z,z')
-\sum_{v}V_{cv\blk}(t)G_{vc'\blk}(z,z')&=\d(z,z')	,
\label{eomGccpk}
\end{align}
\begin{align}
\Big[i\frac{d}{dz}-\e_{v\blk}\Big]G_{vc'\blk}(z,z')
-\sum_{c''}V_{vc''\blk}(t)G_{c''c'\blk}(z,z')&=0	,
\end{align}
\end{subequations}
where  $\e_{\m\blk}\equiv \e^{\rm eq}_{\m\blk}+V_{\m\m\blk}$ are the 
renormalized quasiparticle energies.
The second equation is solved by
\begin{align}
G_{vc'\blk}(z,z')=\int d\bar{z}\;g_{v\blk}(z,\bar{z})
\sum_{c''}V_{vc''\blk}(\bar{t})G_{c''c'\blk}(\bar{z},z'),
\label{solGvc}
\end{align}
where $g_{v\blk}$ is the solution of 
\begin{align}
\Big[i\frac{d}{dz}-\e_{v\blk}\Big]g_{v\blk}(z,z')=\d(z,z')	.
\label{eomgvk}
\end{align}
Inserting Eq.~(\ref{solGvc}) into Eq.~(\ref{eomGccpk}) we find
\begin{align}
\Big[i\frac{d}{dz}-\e_{c\blk}\Big]G_{cc'\blk}(z,z')
-\sum_{c''}\int d\bar{z}\;\S_{cc''\blk}(z,\bar{z})
G_{c''c'\blk}(\bar{z},z')
=\d(z,z')	,
\label{eomGccp2}
\end{align}
where the HSEX self-energy reads
\begin{align}
\S_{cc'\blk}(z,z')&=
\sum_{v}V_{cv\blk}(t) g_{v\blk}(z,z')
V^{\ast}_{c'v\blk}(t').
\label{sigmahsex1}
\end{align}

Equation~(\ref{sigmahsex1}) can be further manipulated. 
The HSEX dynamics of the density matrix is governed by the equation of motion 
$i\frac{d}{dt}\r_{\blk}=[h^{\rm HSEX}_{\blk},\r_{\blk}]$ with 
$h^{\rm 
HSEX}_{\m\n\blk}=\d_{\m\n}\e^{\rm eq}_{\m\blk}+V_{\m\n\blk}$. 
Therefore we have
\begin{align}
i\frac{d}{dt}\r_{cv\blk}&=(\e^{\rm eq}_{c\blk}-\e^{\rm eq}_{v\blk})\r_{cv\blk}+
\sum_{c'}V_{cc'\blk}\r_{c'v\blk}-\sum_{v'}\r_{cv'\blk}V_{v'v\blk}
+
\sum_{v'}V_{cv'\blk}\r_{v'v\blk}-\sum_{c'}\r_{cc'\blk}V_{c'v\blk}
\nn\\
&\simeq
(\e_{c\blk}-\e_{v\blk})\r_{cv\blk}+(f_{v\blk}-f_{c\blk})V_{cv\blk},
\label{eomrfdk}
\end{align}
where we implement the previously introduced approximations  $V_{cc'\blk}\simeq 
\d_{cc'}V_{cc\blk}$, $V_{vv'\blk}\simeq 
\d_{vv'}V_{vv\blk}$, as well as 
$\r_{cc'\blk}\simeq \d_{cc'}f_{c\blk}$, 
$\r_{vv'\blk}\simeq \d_{vv'}f_{v\blk}$, $f_{\m\blk}$ being the 
occupation of the quasiparticle state $\m\blk$.
Taking into account that the off-diagonal 
HSEX potential $V_{cv\blk}(t)$ is given by Eq.~(\ref{hsexpotcv}), 
and comparing with the BSE in Eq.~(\ref{bse1}), we 
see that 
Eq.~(\ref{eomrfdk}) is solved by  
\begin{align}
\r_{cv\blk}(t)=\sum_{\l}A^{\l\bz}_{cv\blk}\r_{\l}e^{-iE_{\l\bz}t}.
\end{align}
Ignoring the finite-density corrections to the 
exciton wavefunctions, we have 
\begin{align}
\sum_{c'v'\blk'}K^{\bz}_{cv\blk,c'v'\blk'}
A^{\l\bz}_{c'v'\blk'}=(\e^{\rm eq}_{c\blk}-\e^{\rm 
eq}_{v\blk}-E_{\l\bz}^{\rm eq})A^{\l\bz}_{cv\blk}=K^{\l\bz}_{cv\blk},
\end{align}
and therefore 
$V_{cv\blk}(t)=\sum_{\l}K^{\l\bz}_{cv\blk}\r_{\l}e^{-iE_{\l\bz}t}$. 
Substituing this result into Eq.~(\ref{sigmahsex1}) we obtain
\begin{align}
\S_{cc'\blk}(z,z')&=
\sum_{v\l\l'}
K^{\l\bz}_{cv\blk}\r_{\l}e^{-iE_{\l\bz}t}
g_{v\blk}(z,z')e^{iE_{\l'\bz}t'}
K^{\l'\bz\ast}_{c'v\blk}\r^{\ast}_{\l'}.
\label{shsexk}
\end{align}

Let us extract the retarded component of the self-energy. From Eq.~(\ref{eomgvk}) we have
\begin{align}
g_{v\blk}^{R}(t,t')=-i\th(t-t')e^{-i\e_{v\blk}(t-t')}.
\end{align}
For sufficiently long probes, beating terms can be 
ignored~\cite{rustagi_coherent_2019,zhang_long-lived_2025}. 
Retaining only terms with $\l=\l'$, the retarded component of 
Eq.~(\ref{shsexk}) in frequency space reads
\begin{align}
\S^{R}_{cc'\blk}(\w)=
\sum_{\l v}
K_{cv\blk}^{\l\bz}\frac{|\r_{\l}|^{2}}{z-E_{\l\bz}-\e_{v\blk}}
K^{\l\bz\ast}_{c'v\blk}-i\eta.
\label{shsexk2}
\end{align}
This formula generalizes Eq.~(\ref{twobself}) to multiple bands.
Accordingly, the generalization of Eq.~(\ref{twobself2}) for the lesser 
self-energy is 
\begin{align}
\S^{<}_{cc'\blk}(\w)&=-\sum_{\l v}
f(\w-E_{\l\bz}-\e_{v\blk})
\Big[
K_{cv\blk}^{\l\bz}\Big(\frac{|\r_{\l}|^{2}}{z-E_{\l\bz}-\e_{v\blk}}
-\frac{|\r_{\l}|^{2}}{z^{\ast}-E_{\l\bz}-\e_{v\blk}}\Big)
K^{\l\bz\ast}_{c'v\blk}
-2i\eta\Big].
\label{twobselfgen}
\end{align}
Equations~(\ref{shsexk2}) and (\ref{twobselfgen}) are the 
self-energies of the main text for the coherent regime, i.e., 
Eqs.~(\ref{sigma}) with $\s$ evaluated in 
$N_{\l\blQ}=\d_{\blQ\bz}|\r_{\l}|^{2}$.

\section{TX self-energy}
\label{txapp}

We expand the xc function in the excitonic 
basis~\cite{perfetto_first-principles_2016,stefanucci_excitonic_2025} 
\begin{align}
L^{\blQ\lessgtr}_{c_{1}v_{1}\blk_{1},c_{2}v_{2}\blk_{2}}(t,t')
=\sum_{\l}A^{\l\blQ}_{c_{1}v_{1}\blk_{1}}
N^{\lessgtr}_{\l\blQ}e^{-iE_{\l\blQ}(t-t')}
A^{\l\blQ\ast}_{c_{2}v_{2}\blk_{2}},
\end{align}
with $N^{<}_{\l\blQ}=N^{\rm inc}_{\l\blQ}$ and 
$N^{>}_{\l\blQ}=1+N^{\rm inc}_{\l\blQ}$, and approximate the 
Green's function at the quasiparticle level, 
$G^{\lessgtr}_{v\blk}(t,t')=if^{\lessgtr}_{v\blk}e^{-i\e_{v\blk}(t-t')}$,
with $f^{<}_{v\blk}=f_{v\blk}$ and $f^{>}_{v\blk}=f_{v\blk}-1$.
Then, the Fourier transform of the lesser and greater TX self-energy, 
see Eq.~(\ref{sigma<>}), 
read
\begin{align}
\S^{\lessgtr}_{cc'\blk}(\w)&=2\p i
\sum_{\l\blQ v}K^{\l\blQ}_{cv\blk}\;N^{\lessgtr}_{\l\blQ}
f^{\lessgtr}_{v\blk-\blQ}\;
K^{\l\blQ\ast}_{c'v\blk}
 \d(\w-E_{\l\blQ}-\e_{v\blk-\blQ}).
\label{sigma3}
\end{align}
The retarded TX self-energy can be calculated from 
\begin{align}
\S^{R}_{cc'\blk}(\w)&=i\int\frac{d\w'}{2\p}
\frac{\S^{>}_{cc'\blk}(\w)-\S^{<}_{cc'\blk}(\w)}{\w-\w'+i\eta}-i\eta
=\sum_{\l\blQ v}K^{\l\blQ}_{cv\blk}\;
\frac{N_{\l\blQ}+1-f_{v\blk-\blQ}}{\w-E_{\l\blQ}-\e_{v\blk-\blQ}+i\eta}\;
K^{\l\blQ\ast}_{c'v\blk}-i\eta.
\label{sigmaRapp}
\end{align}
For not too large excitation densities, we can approximate 
$1-f_{v\blk-\blQ}\simeq 0$, thus recovering the result in the main 
text.
In analogy with Eq.~(\ref{twobselfgen}), we approximate the lesser TX self-energy 
as
\begin{align}
\S^{<}_{cc'\blk}(\w)=- \sum_{\l\blQ v} 
f(\w-E_{\l\blQ}-\e_{v\blk-\blQ})
\Big[K^{\l\blQ}_{cv\blk}\Big(\frac{N_{\l\blQ}}{z-E_{\l\blQ}-\e_{v\blk-\blQ}}
-\frac{N_{\l\blQ}}{z^{\ast}-E_{\l\blQ}-\e_{v\blk-\blQ}}\Big)
K^{\l\blQ\ast}_{c'v\blk}-2i\eta\Big].
\label{twobselfgenic}
\end{align}
Equations~(\ref{sigmaRapp}) and (\ref{twobselfgenic}) are the 
self-energies of the main text for the incoherent regime, i.e., 
Eqs.~(\ref{sigma}) with $\s$ evaluated in 
$N_{\l\blQ}=N_{\l\blQ}^{\rm inc}$.

\section{Numerical details}
\label{numapp}
The electronic structure is described using a spin-resolved density functional theory (DFT) band dispersion
$\varepsilon_{n\mathbf{k}}$, mapped onto a tight-binding representation as detailed in Ref.~\cite{PhysRevB.88.085433}.
To match the experimentally measured quasiparticle band gap of about
$2.0~\mathrm{eV}$, a rigid scissor correction of
$0.5~\mathrm{eV}$ is applied to the conduction-band states $\varepsilon_{c\mathbf{k}}$
\cite{PhysRevLett.113.076802,Hsu_2019}.
The eigenstates of the Bloch Hamiltonian, denoted by $\mathbf{U}_{\mathbf{k}}$,
serve as the basis for constructing the Coulomb matrix elements that appear in both the
Bethe–Salpeter equation (BSE) kernel and the Hartree–screened-exchange (HSEX) Hamiltonian,
following the methodology of Ref.~\cite{perfetto_real-time_2023}.

The Coulomb interaction is modeled using the Rytova–Keldysh potential
\cite{keldysh,PhysRevB.84.085406}, assuming a dielectric constant $\varepsilon = 1$
and a screening length of
$r_0 = 45~\mathring{\rm A}$ \cite{stier_magnetooptics_2018}.
To treat the divergence of the Coulomb potential $v_{\mathbf{q}}$ near the $\Gamma$ point,
we adopt the regularization scheme introduced in Refs.~\cite{PhysRevB.88.245309,PhysRevB.97.205409},
which consists of averaging $v_{\mathbf{q}}$ over a small region $\Omega$ around $\mathbf{q}=0$.
The size of this region is determined by the spacing of the Brillouin-zone sampling grid.

In typical semiconductors, Coulomb matrix elements that involve processes changing the number of electrons
in the valence or conduction bands are vanishingly small and are therefore neglected
\cite{PhysRevB.93.205145}.
Since the Rytova–Keldysh interaction tensor $V$ in Eq.~(5) already incorporates
\emph{ab initio} static screening over the momentum scales relevant for excitonic physics,
the screened exchange interaction $W$ entering both the BSE kernel and the HSEX Hamiltonian
is approximated by $W \approx V$.
This simplification has been demonstrated to reproduce experimental exciton binding energies
with high accuracy in two-dimensional systems
\cite{steinhoff2,PhysRevLett.113.076802,PhysRevB.88.045318}.

In the calculations presented here, the electronic subspace is limited to the two uppermost valence bands
and the two lowest conduction bands.
The Brillouin zone is sampled using a uniform mesh comprising $3072$ $\mathbf{k}$ points.

\end{widetext}

\end{document}